# Nanodroplets Behavior on Graphdiyne Membranes

Ygor M. Jaques and Douglas S. Galvão
Applied Physics Department, University of Campinas, Campinas, SP 13081-970, Brazil

## ABSTRACT

In this work we have investigated, by fully atomistic reactive (force field ReaxFF) molecular dynamics simulations, some aspects of impact dynamics of water nanodroplets on graphdiyne-like membranes. We simulated graphdiyne-supported membranes impacted by nanodroplets at different velocities (from 100 up to 1500 m/s). The results show that due to the graphdiyne porous and elastic structure, the droplets present an impact dynamics very complex in relation to the ones observed for graphene membranes. Under impact the droplets spread over the surface with a maximum contact radius proportional to the impact velocity. Depending on the energy impact value, a number of water molecules were able to percolate the nanopore sheets. However, even in these cases the droplet shape is preserved and the main differences between the different impact velocities cases reside on the splashing pattern at the maximum spreading.

## INTRODUCTION

The materials science revolution created by the advent of graphene [1] has renewed the interest for other two-dimensional carbon allotropes, such as graphynes [2–5]. Graphyne is a generic name for a family of 2D carbon allotropes where acetylenic groups connect benzenoid-like rings, with the coexistence of sp and sp2 hybridized carbon atoms. Similarly to graphene, tubular structures can also exist [3,4]. Graphdiynes are graphyne-like structures with double acetylene linkages (Figure 1). Graphdiynes can exist in many different configurations, the most common are known as α, β, and γ-graphdiynes (see Figure 1 for an example of γ-graphdiyne). They constitute a family of interesting class of membranes with uniformly distributed nanopores [5–7] that can be exploited for a variety of applications. They are one of the most stable non-natural carbon allotropes and some structures have been already experimentally realized [6].

In order to study the applicability of graphdiynes as selective membranes [8,9], it is important to known their wettability behavior on equilibrium and also on non-equilibrium conditions, such as like under high velocity impacts. Droplet splashing dynamics on surfaces is an important area in general science and industry [10–13], dealing with diverse aspects that a liquid could have (such as shapes and sizes) on impacting surfaces at different velocities. In this work, we have use fully atomistic molecular dynamics (MD) simulations to investiagate how water nanodroplets behave on impact at graphdyine membranes. Due to space limitations we restrict ourselves to the case of γ-graphdyines.

## THEORY

The MD simulations were carried out using the reactive force field ReaxFF [14], as implemented on LAMMPS package [15]. We considered structures (Figure 1) consisting of γ-graphdiyne membranes of 100 x 100 Å² and droplets with radii values of about 20 Å (approximately 3000 water molecules).

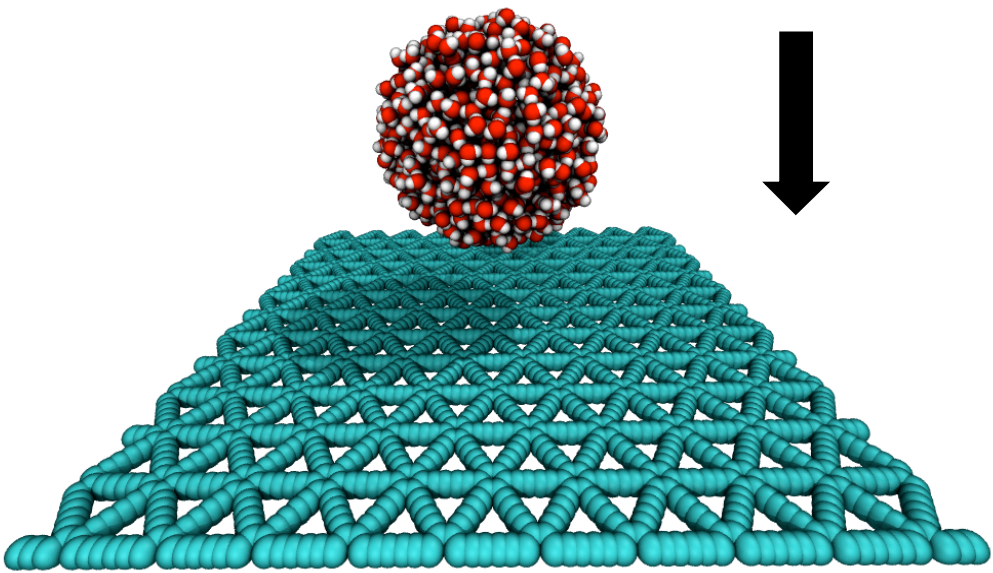

**Figure 1.** Initial set up configuration for water nanodroplets impacting on γ-graphdyine membranes.

Before running the impact MD simulations, we equilibrated the water droplets using a NVT ensemble coupled to a Nosé-Hoover thermostat [16,17] to keep the temperature (300 K) constant. After that, velocities ranging from 100 to 1500 m/s downwards were added to the droplets, shooting them against the graphdiyne membranes (Figure 1).
The impact MD simulations were carried out using a NVE ensemble, with time steps of 0.1 fs. We chose to use a reactive force field (ReaxFF) because of the possible large structural deformations, which are not well described for non-reactive force fields (most of them use quadratic approximations). Membrane borders were kept fixed to best mimic a supporting substrate.

**DISCUSSION**

Our results show that water droplets impacting on γ-graphdiyne membranes exhibit splashing behaviors similar to the ones observed for graphene membranes [18] only for small velocities values. The graphdiyne higher porosity and flexibility (in relation to graphene) create a very complex droplet dynamics, which depending on the impact velocity values can result in partial liquid membrane perfusion,
In Figure 2 we present a representative MD snapshot of the equilibrium configuration for the case of impact velocity of 100 m/s. Due to the graphdiyne high porosity and flexibility, a small deformation was observed around the region of droplet contact. Although some water molecules passed through the graphdyine pores at the impact moment, when the final droplet configuration is reached, this no longer occurs. Splashing patterns and droplet equilibrium configurations became more asymmetric increasing the droplet impact to 500 m/s. This can be better evidenced analyzing the droplet surface density profile (Figure 3).

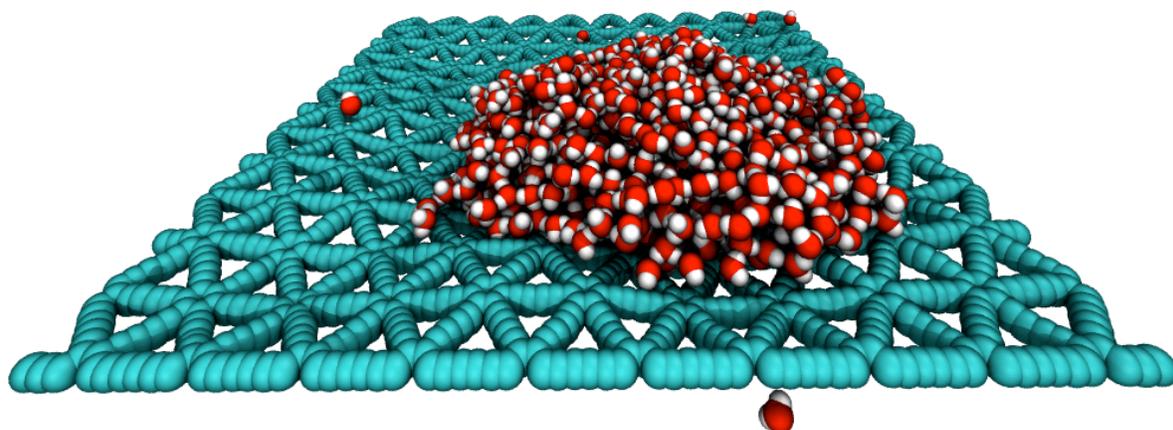

**Figure 2.** MD snapshot of the droplet equilibrium configuration, for the case of 100 m/s impact velocity.

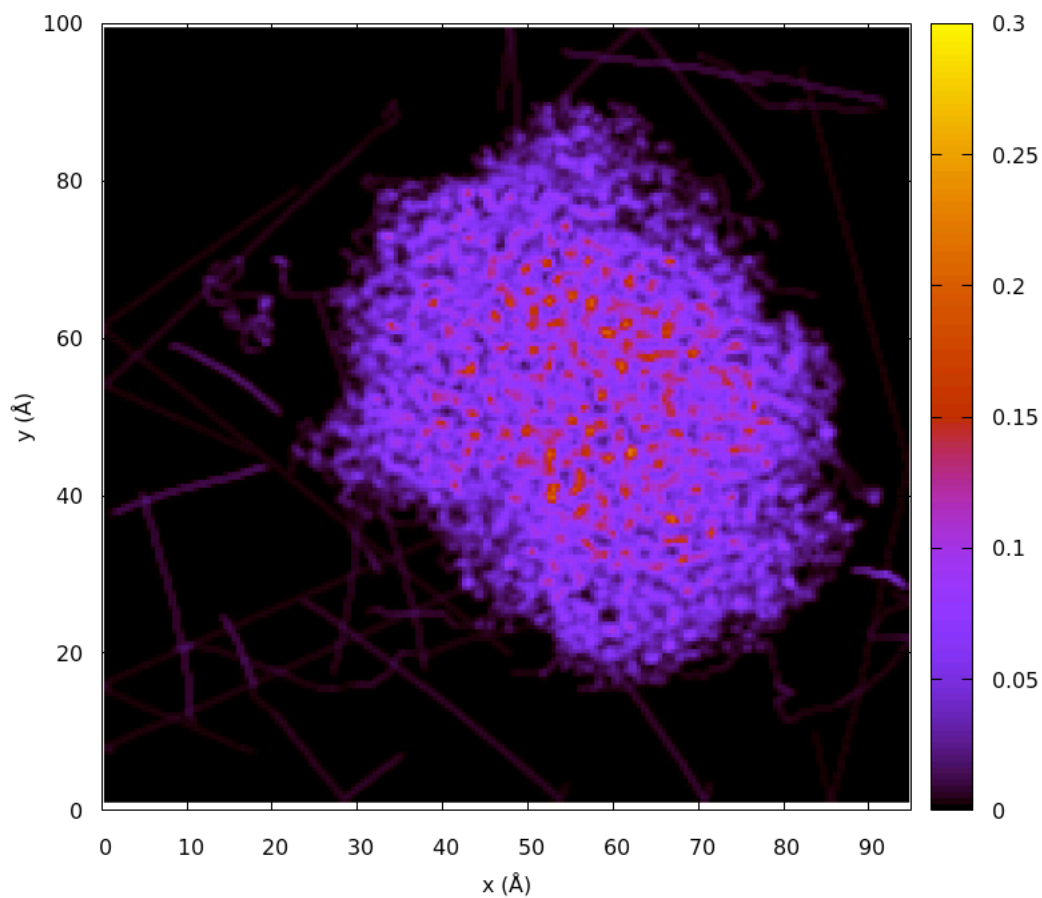

**Figure 3.** Superficial density (arbitrary units) of water molecules at the instant of maximum spreading on the graphdiyne surface. Results for the velocity case of 500 m/s.

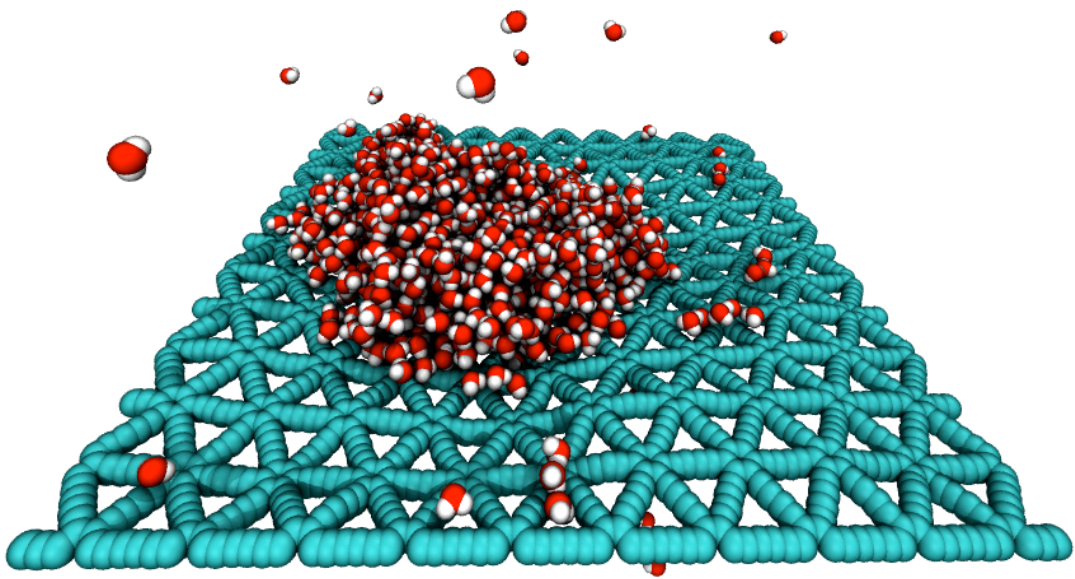

**Figure 4.** MD snapshot of the equilibrium configuration, for the case of 1000 m/s impact velocity.

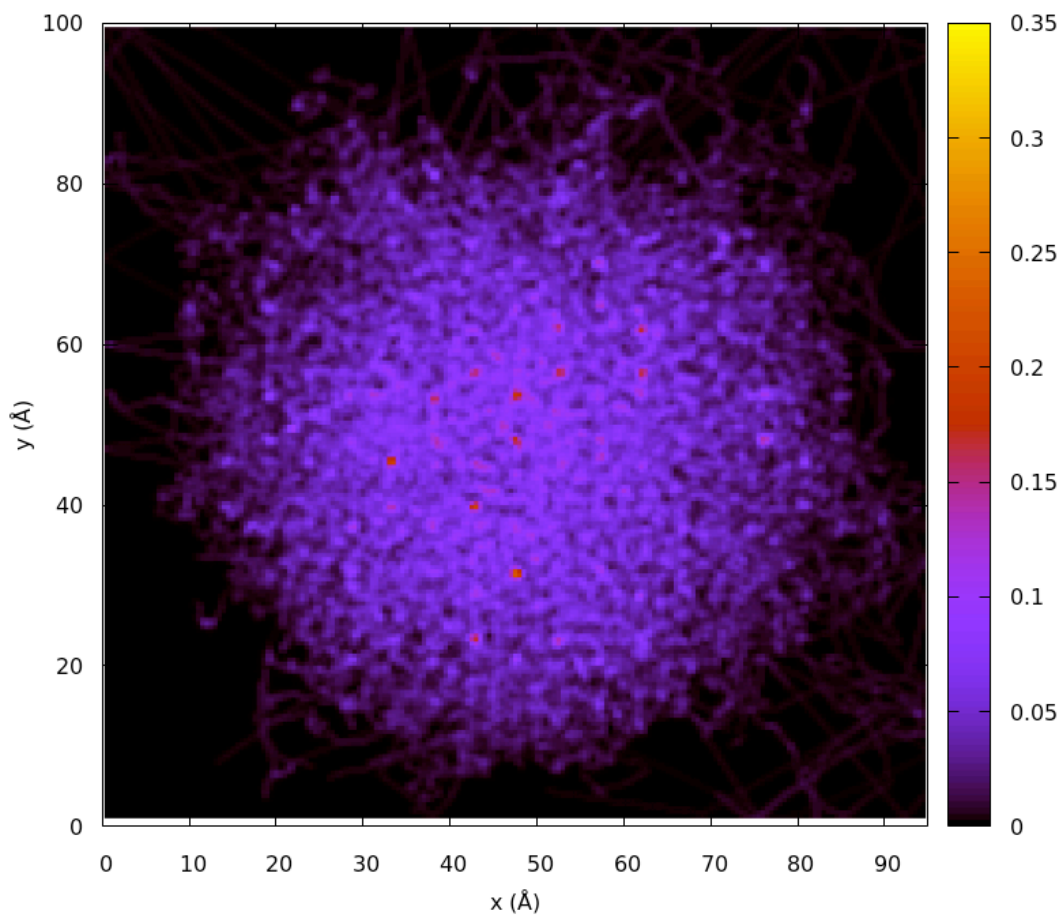

**Figure 5.** Superficial density (arbitrary units) of water molecules at the instant of maximum spreading along graphdiyne surface. Results for the velocity case of 1500 m/s.

The density map shows that at this impact velocity value (500 m/s), the largest water molecules density (at maximum spreading) remains at the nanodroplet center. In our previous study with graphene sheets, the density of water molecules was more homogeneous at maximum spreading at this velocity. Also, for graphene, the droplet shape only became asymmetric for higher impact velocities (750 m/s). For graphdiyne membranes, we can attribute this droplet deformation due to fact that graphdiyne it much more flexible than graphene and also to its intrinsic structural asymmetry.

Similarly, to the observed for the case of 100 m/s, some water molecules can pass through the graphdiyne membrane, but the majority of the liquid retracts after impact and stays in a stable configuration. For the droplet impact velocity of 1000 m/s, the splashing patterns continue to show the same asymmetric shapes at maximum spreading (Figure 4).

However, for the largest velocity value we have considered in this study (1500 m/s), the patterns at maximum spreading change again, becoming more symmetric and more homogeneous. The spherical-like distribution of water molecules on over all droplet volume is recovered (Figure 5). Interestingly, the number of water molecules passing through the pores changes only slightly in relation to the case of 1000 m/s.

These results suggest that exists a velocity threshold determining the symmetry level of the spreading droplet. Near the droplet fragmentation point, where the droplet will not retract anymore, the spreading pattern changes again from asymmetric shapes to more symmetrical (spherical) ones. Preliminary analysis seems to indicate that this is associated with the level of the membrane structural (bending) deformations. Further investigations are needed to confirm the existence of a relationship between membrane bending and the symmetry of the splashing patterns.

**CONCLUSIONS**

We have investigated through fully atomistic reactive molecular dynamics simulations the impact dynamics of water nanodroplets on graphdiyne membranes. The graphdiyne higher porosity and flexibility in relation to graphene create a very complex droplet dynamics. Depending on the impact velocity, partial liquid membrane perfusion can occur.

For small impact velocity values, there is a good agreement between the micro and nanoscales, with the water droplet having symmetric splashing patterns on graphdiyne membranes. However, as we increase the impact velocity values (in our system this range was 500-1000 m/s) the splashing patterns become irregular and asymmetric, but still the majority of water molecules stay at the center of droplet on maximum droplet spreading. If we continue to increase the impact velocity value to1500 m/s, the spreading patterns change again, the splashing recover the symmetrical shapes and there is a more homogeneous distribution of water molecules on over all droplet volume.

Preliminary analysis seems to indicate that the symmetry changes in spreading patterns and droplet surface density values are related to the level of the maximum graphdiyne membrane structural (bending) deformations upon droplet impacts. Further investigations are needed to confirm the existence of this relationship.

ACKNOWLEDGMENTS

This work was supported in part by the Brazilian Agencies CAPES, CNPq and FAPESP. The authors thank the Center for Computational Engineering and Sciences at Unicamp for financial support through the FAPESP/CEPID Grant # 2013/08293-7.

**REFERENCES**

1. K. S. Novoselov et. al., Science **306**, 666 (2004).
2. R. H. Baughman, H. Eckhardt, and M. Kertesz, J. Chem. Phys. **87**, 6687 (1987).
3. V. R. Coluci, S. F. Braga, S. B. Legoas, D. S. Galvao, and R. H. Baughman, Phys. Rev. B **68**, 35430 (2003).
4. V. R. Coluci, S. F. Braga, S. B. Legoas, D. S. Galvao, and R. H. Baughman, Nanotechnology **15**, S142 (2004).
5. P. A. S. Autreto, J. M. de Sousa, and D. S. Galvao, Carbon N. Y. **77**, 829 (2014).
6. G. X. Li, Y. L. Li, H. B. Liu, Y. B. Guo, Y. J. Li, and D. B. Zhu, Chem Commun **46**, 3256 (2010).
7. Y. Jiao, A. Du, M. Hankel, Z. Zhu, V. Rudolph, and S. C. Smith, Chem. Commun. **47**, 11843 (2011).
8. X. Gao, J. Zhou, R. Du, Z. Xie, S. Deng, R. Liu, Z. Liu, and J. Zhang, Adv. Mater. **28**, 168 (2016).
9. S. Lin and M. J. Buehler, Nanoscale **5**, 11801 (2013).
10. A. L. Yarin, Annu. Rev. Fluid Mech. **38**, 159 (2006).
11. G. Juarez, T. Gastopoulos, Y. Zhang, M. L. Siegel, and P. E. Arratiab, Phys. Fluids **24**, 2012 (2012).
12. R. F. Allen, J. Colloid Interface Sci. **51**, 350 (1975).
13. J. Liu, H. Vu, S. S. Yoon, R. a. Jepsen, and G. Aguilar, At. Sprays **20**, 297 (2010).
14. K. Chenoweth, A. C. T. van Duin, and W. a Goddard, J. Phys. Chem. A **112**, 1040 (2008).
15. S. Plimpton, J. Comput. Phys. **117**, 1 (1995).
16. S. Nosé, J. Chem. Phys. **81**, 511 (1984).
17. W. G. Hoover, Phys. Rev. A **31**, 1695 (1985).
18. Y. M. Jaques, G. Brunetto, and D. S. Galvão, MRS Adv. **1**, 675 (2016).